# THE ROLE OF SURFACE DEFECTS IN THE ADSORPTION OF METHANOL ON $Fe_3O_4$(001)

**Oscar Gamba[1], Jan Hulva[1], Jiri Pavelec[1], Roland Bliem[1], Michael Schmid[1], Ulrike Diebold[1], Gareth S. Parkinson[1,*].**

[1]Institute of Applied Physics, TU Wien, Vienna, Austria.

*parkinson@iap.tuwien.ac.at

**ABSTRACT**

The adsorption of methanol ($CH_3OH$) at the $Fe_3O_4$(001)-($\sqrt{2}\times\sqrt{2}$)R45° surface was studied using X-ray photoelectron spectroscopy (XPS), scanning tunneling microscopy (STM), and temperature-programmed desorption (TPD). $CH_3OH$ adsorbs exclusively at surface defects sites at room temperature to form hydroxyl groups and methoxy ($CH_3O$) species. Active sites are identified as step edges, iron adatoms, antiphase domain boundaries in the ($\sqrt{2}\times\sqrt{2}$)R45° reconstruction, and above Fe atoms incorporated in the subsurface. In TPD, recombinative desorption is observed around 300 K, and a disproportionation reaction to form methanol and formaldehyde occurs at 470 K.

# 1. INTRODUCTION

Methanol ($CH_3OH$), the simplest alcohol, can be involved in several processes to produce hydrogen (oxidative reforming, decomposition, steam reforming) [1] and has received renewed interest for its importance in fuel-cell technology [2]. Moreover, methanol chemisorption has been termed a "smart chemical probe" [3] to study active sites on metal oxide catalysts because adsorption allows to quantify the density of active sites, while the product distribution observed upon desorption is thought to reflect the nature of the active sites.

Studies of methanol adsorption on well-characterized metal-oxide surfaces have sought to correlate the atomic-scale structure with chemical reactivity [3-13]. Methanol does not typically chemisorb at bulk truncated oxide surfaces, and some degree of additional coordinative unsaturation is required for dissociative adsorption, for example at step edges [11, 12]. Oxygen vacancies ($V_O$) have been shown to be the major active sites on $TiO_2$ [7, 8] and $CeO_2$ [9, 10] surfaces, with adsorbed methoxy species ($CH_3O^-$) and hydroxyl groups formed at room temperature. Reaction products such as formaldehyde and methane are reported to emerge from this chemistry. However, as noted by Vohs in his recent review of oxygenate adsorption on metal oxides [4], little is known about the reactivity of isolated cation defects.

Iron oxides represent an interesting case in this regard because their bulk defect chemistry is dominated by the cation sublattice, with little evidence that $V_O$s form in the bulk, or at the surface. Indeed, a recent study on the $Fe_3O_4(111)$ surface [14] found dissociative adsorption on Fe-terminated regions of the surface, and concluded this to occur via a Brønsted acid-base mechanism requiring undercoordinated cation–anion pairs. Recombination to produce methanol was observed at 330-360 K in TPD, along with a disproportionation reaction between two adsorbed methoxy species to produce methanol and formaldehyde as follows [14]

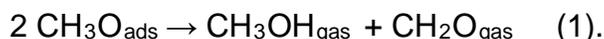
$$2\ CH_3O_{ads} \rightarrow CH_3OH_{gas} + CH_2O_{gas} \quad (1).$$

In this paper, we study the adsorption of $CH_3OH$ on the magnetite $Fe_3O_4$(001) surface, with an emphasis on the role of isolated cation defects. This surface exhibits a $(\sqrt{2}\times\sqrt{2})R45°$ reconstruction over a wide range of oxygen chemical potentials [15] that is based on a rearrangement of the cations in the second and third surface layers [16]. The reconstructed unit cell, indicated by a purple square in Figure 1a, contains four octahedrally coordinated $Fe_{oct}$ atoms (dark blue spheres) and eight oxygen atoms (red spheres) in the surface layer. The second layer contains three tetrahedrally coordinated $Fe_{tet}$ atoms (light blue spheres), one of which is an additional interstitial atom labelled $Fe_{int}$. Essentially, this atom replaces two $Fe_{oct}$ atoms in the third layer resulting in a more oxidized surface. According to angle resolved XPS and DFT+U calculations, all Fe atoms in the outermost four layers are $Fe^{3+}$ [16].

In prior studies, formic acid and, to some degree, water have been found to dissociate on this surface at room temperature [17, 18]. Here, using a combination of STM, XPS, and TPD, we demonstrate that methanol adsorption is restricted to surface defects. Specifically, the active sites are determined to be step edges, Fe adatoms, antiphase domain boundaries (APDBs) in the $(\sqrt{2}\times\sqrt{2})R45°$ reconstruction, and Fe atoms incorporated in the subsurface. Adsorption at the former two defects is due to the high coordinative unsaturation of the cations at such sites, whereas reactivity at the latter two defects is linked to the presence of $Fe^{2+}$. Desorption occurs via two channels in TPD; recombination to methanol, and a disproportionation reaction to form methanol and formaldehyde.

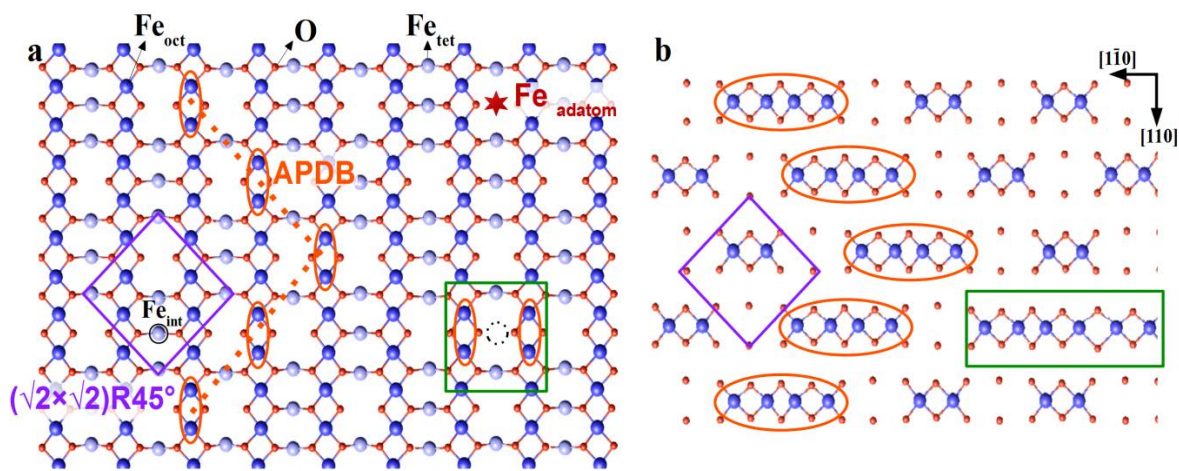

**Figure 1. Defects at the Fe$_3$O$_4$(001)-(√2×√2)R45° surface**. (a) The outermost two layers of the Fe$_3$O$_4$(001)-(√2×√2)R45° surface are shown in panel (a) and the third layer in panel (b). The (√2×√2)R45° unit cell, indicated by the purple square, contains four Fe$_{oct}$ atoms (big, balls, dark blue) and eight O atoms (small balls, red) in the surface layer, and three Fe$_{tet}$ atoms in the second layer (big balls, light blue). The Fe$_{tet}$ indicated by the black circle is an interstitial (Fe$_{int}$) linked to the subsurface cation vacancy (SCV) reconstruction [16]. The interstitial replaces two Fe$_{oct}$ atoms in the third layer, such that there is only two Fe$_{oct}$ per unit cell instead of four. The dotted orange line indicates an antiphase domain boundary (APDB) in the SCV reconstruction. The pairs of surface Fe$_{oct}$ atoms indicated by orange ovals at the boundary appear bright in STM images. Note that four Fe$_{oct}$ are present in a row in the third layer beneath the APDB (see panel b). The green boxes highlight a defect in which an additional Fe$_{oct}$ atom is incorporated in the third layer, which causes Fe$_{int}$ to relocate to the other Fe$_{oct}$ vacancy position. Note that because six Fe$_{oct}$ are present in a row in the third layer beneath this defect, it appears as a double protrusion in STM images (orange

ovals). The black dashed circle represent the position for a missing $Fe_{int}$ in the second layer. The red star represents a possible position of an Fe adatom.

## 2. MATERIALS AND METHODS

The experiments were performed in ultrahigh vacuum (UHV). A natural $Fe_3O_4$ (001) sample (SurfaceNet GmbH) was prepared *in-situ* by 1 keV $Ar^+$ sputtering at room temperature for 20 minutes followed by annealing in UHV at 873 K for 15 minutes. Once no contamination could be detected by XPS in the C1*s* region, the sample was annealed in $O_2$ ($5 \times 10^{-7}$ mbar) at 873 K for 15 minutes, which results in a surface with the SCV reconstruction. To create a surface with an increased defect density, Fe was evaporated from a 2-mm-thick rod (99.99 + %, MaTeck GmbH) at room temperature using an Omicron electron-beam evaporator; the deposition rate was calibrated by a quartz crystal microbalance. Methanol was obtained from Sigma Aldrich at a purity of 99.8% and purified with several freeze-pump-thaw cycles. For the STM experiments methanol vapour was dosed into the background in the chamber through a high-precision leak valve.

STM measurements were performed at room temperature using an Omicron UHV-STM-1 instrument in constant current mode with electrochemically etched tungsten tips. The base pressure was below $10^{-10}$ mbar.

XPS and TPD measurements were performed in a second vacuum system (base pressure $5 \times 10^{-11}$ mbar). XPS spectra were measured using a SPECS FOCUS 500 monochromatic source (Al Kα) and a SPECS PHOIBOS 150 electron analyzer at normal emission with a pass energy of 16 eV. TPD experiments were performed using a HIDEN HAL/3F RC 301 PIC quadrupole mass spectrometer (QMS). The sample was cooled by a Janis ST-400 UHV liquid-He flow cryostat, and heated by direct current at a rate of 1 K/s through a Ta back plate, on which the sample was mounted. The temperature was measured by a K-type thermocouple, and the sample was biased at -100 V during TPD measurements to prevent electrons from the QMS filament from reaching the sample. For the TPD and XPS measurements, methanol was dosed using a home-built effusive molecular beam source, which enables precise and reproducible exposures to a defined area on the sample surface [19].

## 3. RESULTS

3.1. Defects on the as-prepared surface

Figure 2a shows an STM image of the as-prepared $Fe_3O_4$(001) surface. Rows of protrusions separated by 5.9 Å are due to fivefold coordinated surface iron atoms (the dark blue balls in Figure 1a) within the subsurface cation vacancy (SCV) reconstruction [16]. Surface oxygen atoms (red in Figure 1a) are not imaged as there are no O-derived states in the vicinity of the Fermi level [16]. A step edge runs across the centre of the image from left to right, separating two adjacent terraces (yellow arrow). The apparent step height of 2.1 Å corresponds to the spacing between equivalent planes in the bulk structure [20] (i.e., the first and third layer in Figure 1). Note that the direction of the iron rows rotates 90° when going from one terrace to the next [20], consistent with the inverse spinel structure of magnetite. As reported previously [20], steps that run parallel to the Fe rows on the upper terrace are generally straight, whereas perpendicular steps are often jagged.

A second, extended defect that is frequently observed on the freshly prepared surface is the antiphase domain boundary (APDB) [21], indicated by orange arrows in Figure 2a. This feature appears as a chain of bright protrusions located on the $Fe_{oct}$ rows, and is typically aligned at 45° with respect to the row direction (see also Figure 1a) [21]. The APDBs probably arise because the ($\sqrt{2}\times\sqrt{2}$)R45° reconstruction is lifted during each annealing cycle [22], and then renucleates on cooling through 723 K with one of two distinct registries with respect to the underlying bulk. It was noted previously [21] that the APDB forms such that two "narrow" sites of the reconstruction meet at the interface. In the light of the SCV reconstruction [16], this preference can be reinterpreted as four $Fe_{oct}$ atoms in a row in the third layer (as illustrated in Figure 1b). The alternative, where two domains would meet with four Fe vacancies would create twofold coordinated O atoms, is expected to be unfavourable. Interestingly, with no $Fe_{int}$ in the second layer and four $Fe_{oct}$ atoms in a row in the third layer, the local structure at the APDB is akin to a bulk truncated surface.

In addition to the line defects, two types of point defects are observed (Figure 2a). Surface hydroxyl groups appear in STM as bright protrusions located on the Fe rows (cyan box). These species were initially identified through the adsorption of atomic H on this surface [23], and have also been observed following dissociative adsorption of water [21]. They are easily distinguished from other defects as they exhibit a characteristic hopping between opposite Fe rows in STM movies collected at room temperature [18, 21, 23]. It is important to note that the OH group is a H atom adsorbed on a surface O atom, albeit it appears as increased brightness of a pair on ajacent Fe atoms. This is an electronic effect, as the OH donates charge to the neighboring Fe atoms, which makes them brighter in STM [18]. Finally, the green boxes highlight pairs of bright features located on neighboring Fe rows. At first glance these features appear similar to hydroxyl groups, but they do have a different apparent height (50 pm, compared to 20 pm for the OH) and they do not exhibit the characteristic hopping behavior described above. In the next section, we demonstrate that these defects are linked to Fe incorporated in the subsurface.

When the surface shown in Figure 2a is exposed to 20 L of $CH_3OH$ at room temperature new features appear at some of the defects. The defects do not change their positions, but the apparent height of some defects increases significantly. For example, the bright features located on neighboring iron rows are now 150 pm high, as compared to the 50 pm previously, see also the line profiles in Figure 3, below. Interestingly, the methanol-induced, bright features were sometimes observed to disappear as the surface was scanned with the STM tip at room temperature, and the defects assumed their original appearance. However, because methanol was still present in the residual gas following the initial exposure, re-adsorption at defects was also observed.

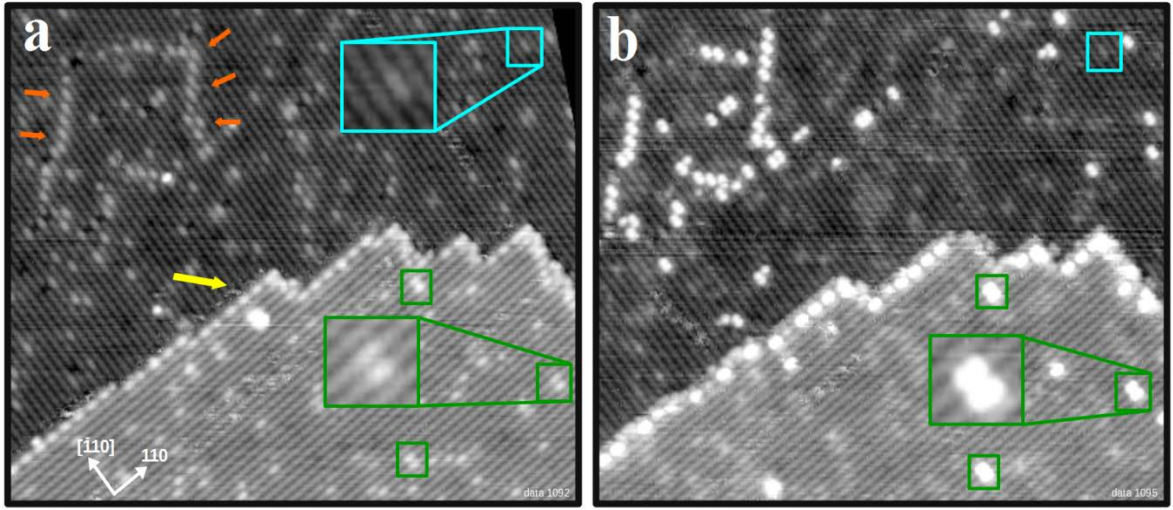

**Figure 2. STM images of the same spot on the Fe$_3$O$_4$ (001) surface before and after exposure to CH$_3$OH at room temperature**. (a) The as-prepared Fe$_3$O$_4$ (001)-($\sqrt{2}\times\sqrt{2}$)R45° surface (39 × 31 nm$^2$; V$_{sample}$ = +1.7 V; I$_{tunnel}$ = 0.3 nA) with typical defects labeled as follows (see also Fig. 1 for a schematic): An APDB is visible as a row of protrusions in the lower terrace (indicated by the orange arrows). The bright protrusion on the Fe rows highlighted by the cyan box is due to a surface hydroxyl. The green boxes highlight pairs of bright features located on neighboring Fe rows, attributed to additional subsurface Fe. (b) Exposure to 20 L CH$_3$OH at room temperature results in bright protrusions located on defects.

3.2. Fe$_3$O$_4$(001)-($\sqrt{2}\times\sqrt{2}$)R45° with additional cation defects

To create an increased coverage of defects, 0.1 monolayer (ML) of Fe was deposited on the as-prepared SCV surface at room temperature, see Figure 3a. (Here 1 ML is defined as one atom per ($\sqrt{2}\times\sqrt{2}$)R45° unit cell, i.e., 1.42 × 10$^{14}$ atoms/cm$^2$). This procedure results in Fe adatoms, as observed previously [24], which appear as bright protrusions between the Fe rows (red circles, see also the red star in the schematic in Fig. 1). The formation of stable adatoms is a distinctive property of the Fe$_3$O$_4$(001)-($\sqrt{2}\times\sqrt{2}$)R45° surface and has been observed for many different elements [refs. 25-29]. The appearance of the Fe protrusions is similar to those observed for adatoms of Au, Ag and Pd [25-27]. In addition, the density of the double-bright features located on neighboring iron rows increases significantly upon Fe deposition (green circles). Similar features have been observed following the deposition of Ni, Co, Ti and Zr [28]. It is known that these elements enter the subsurface, filling one Fe$_{oct}$ vacancy in the third layer of the SCV reconstruction, which induces the Fe$_{int}$ to move and occupy the other. At 1 ML coverage this results in a (1×1) symmetry. By analogy, it is natural to propose that deposited Fe atoms also enter the surface, and locally lift the SCV reconstrucion. In the case of Ni, the bright protrusion associated with the defect was found to appear above the two third-layer Fe$_{oct}$ atoms, rather than above the incorporated foreign metal cation [28]. In the case of Fe incorporation, two such sites are created resulting in two bright protrusions. Note that six Fe$_{oct}$ are present in the third layer at such defects (green boxes in Figure 1).

Exposing the surface shown in Figure 3a to 20 L CH$_3$OH (Figure 3b) leads to similar features as already shown in Figure 2b. Adsorption occurs again at the step edges, APDBs, and the incorporated Fe defects. Frequently after methanol adsorption four distinct maxima are located at each incorporated Fe defect, two protrusions above each Fe$_{oct}$ row (see zoomed areas and line scans in Fig. 3). This suggests that all four Fe$_{oct}$ atoms affected by the subsurface modification can adsorb a methanol related species. Along the row the separation of the maxima is 3 Å (see line scan in Fig. 3), which is consistent with the separation of neighboring $Fe_{oct}$ atoms. The adsorption of methoxy groups in close proximity of at this defect could promote the disproportionation reaction (equation 1).

In addition, fuzzy features appear at the position of the Fe adatoms (red circle). Such apparently noisy parts in STM images are typically associated with weakly bound adsorbates that move during the scan. As before, dynamic desorption and readsorption was observed while scanning the surface with the STM at room temperature.

As mentioned above, there are two kinds of step edges on the $Fe_3O_4$(001) surface. The yellow arrow in Figure 3a highlights a straight step edge, which runs parallel to the iron rows on the upper terrace. The second type, perpendicular to the octahedral iron rows, is more jagged (red arrow).

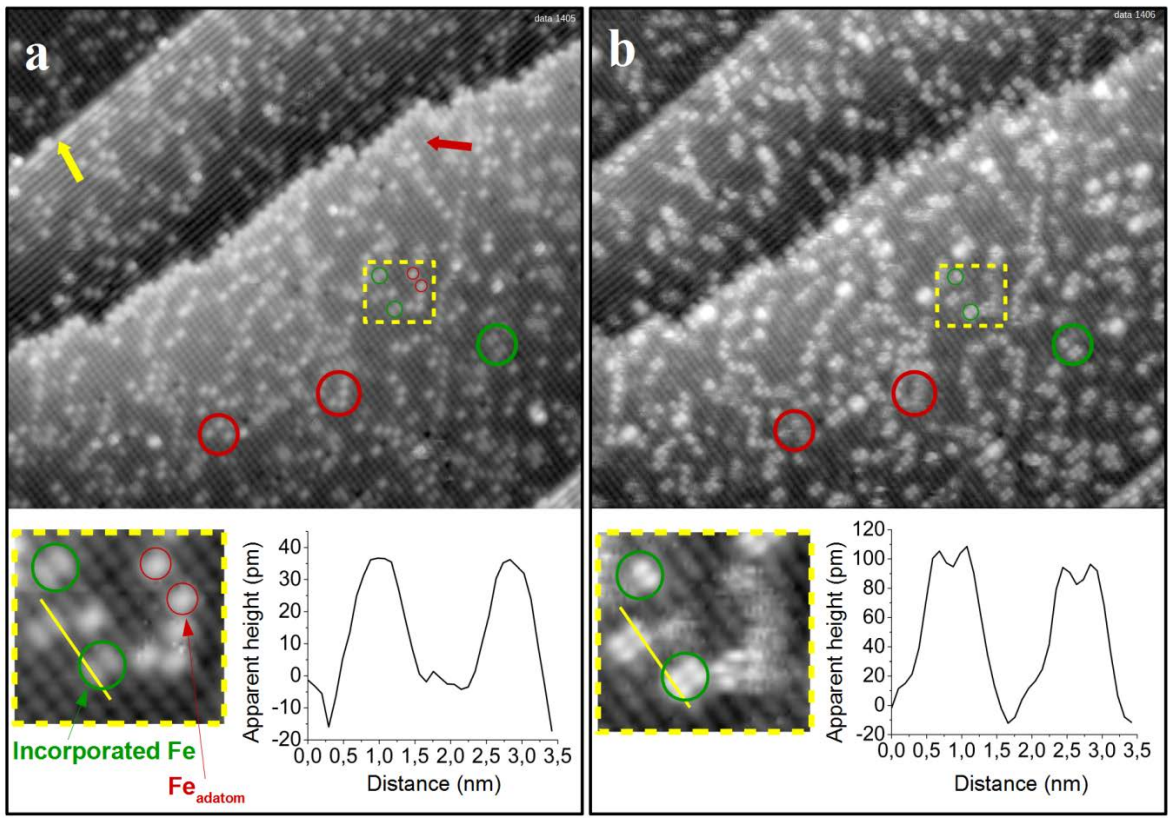

**Figure 3. STM images of the same area of a Fe-rich $Fe_3O_4$ (001) surface before and after exposure to $CH_3OH$ at room temperature.** (a) The as-prepared $Fe_3O_4(001)$-($\sqrt{2}\times\sqrt{2}$)R45° surface modified by deposition of 0.1 ML Fe (48 × 45 nm$^2$; $V_{sample}$ = +1.7 V; $I_{tunnel}$ = 0.3 nA). Fe adsorbs as adatoms (red circles) and incorporates in the subsurface, forming bright double features on the $Fe_{oct}$ rows (green circles), see inset. (b) After exposure to $CH_3OH$, defects have a larger apparent height (see line profiles); the fuzzy appearance of the species on top of Fe adatoms indicate weakly adsorbed molecules.

3.3. XPS

To investigate the nature of the observed protrusions we performed XPS experiments. Figure 4 shows O1$s$ and C1$s$ photoemission spectra that were recorded after the as-prepared $Fe_3O_4(001)$ surface was exposed to 1.8 L of CH$_3$OH at 65 K, and subsequently annealed to progressively higher temperatures. The O1*s* spectra are shown in Figure 4a. The clean Fe$_3$O$_4$(001) surface exhibits a slightly asymmetric peak at 530.1 eV due to the lattice oxygen in magnetite as reported previously [23, 29]. Adsorption of 1.8 L CH$_3$OH at 65 K and annealing to 95 K produces two additional signals. The shoulder on the high-energy side at approximately 531.3 eV is consistent with both surface OH groups and methoxy species [30], while the peak at 532.9 eV is attributed to molecular CH$_3$OH [31]. Neither the shape nor position of the lattice oxygen peak is affected by methanol adsorption, only its intensity is reduced. As the sample is heated to progressively higher temperatures the molecular methanol desorbs first. By 280 K, the intensity of the peak at 532.9 eV has decreased from initially 32 % to just 4 %. The signal from the lattice oxygen increases again.

A similar conclusion can be drawn from the C1s spectra shown in Figure 4b. No detectable C peak is present when the surface is freshly prepared, but following the adsorption of methanol at 65 K and annealing to 95 K a symmetric peak centered at 286.5 eV appears. Upon heating this peak becomes narrower and decreases in intensity. When the sample is heated to 280 K, its intensity decreases notably and its position shifts to 286 eV. This suggests that molecular methanol desorbs below room temperature, leaving only adsorbed methoxy species at 280 K. Methoxy generally has a lower C *1s* binding energy than methanol (e.g. on ZnO [32], TiO$_2$ (110) [8], TiO$_2$ (001) [33], MgO [34], and CeO$_2$ (111) [31]) due to an increase in the electron density around the C atom when the hydroxyl proton is removed [32].

In order to quantitatively determine the methoxy coverage at 300 K from the XPS data we compared the C1s peak area shown in Figure 4b to that of the same surface exposed to a saturation coverage of formic acid (not shown). We have previously shown that formic acid exposure results in a complete monolayer of formate with a density of 2.84 × $10^{14}$ molecules/cm$^2$, or 2 molecules per ($\sqrt{2} \times \sqrt{2}$)R45° unit cell. Since both molecules contain just one C atom, the formate:methoxy ratio of 1:0.0283 suggests a methoxy coverage of 8,05 × $10^{12}$ molecules/cm$^2$. Such a low coverage is consistent with the defect only adsorption observed by STM in Figure 2.

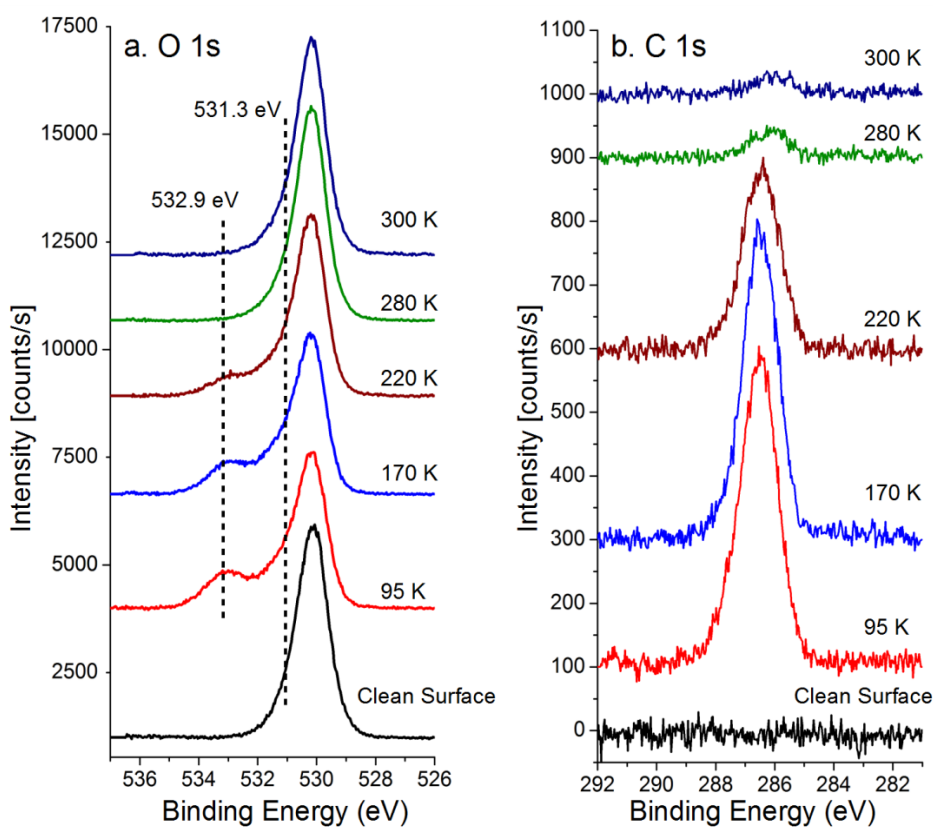

**Figure 4.** C1s and O1s XPS spectra from methanol adsorbed on the Fe$_3$O$_4$(001)-($\sqrt{2} \times \sqrt{2}$)R45° surface at 65 K and annealed as indicated. The data are offset in the y-direction for clarity.

In keeping with the STM experiments, we also performed XPS measurements for a surface on which the defect concentration was enhanced by deposition of 0.3 ML Fe. Figure 5a compares the Fe$2p$ spectra obtained before and after the deposition of the Fe (no methanol exposure). As shown previously [24], the deposition of Fe on the Fe$_3$O$_4$(001) surface results in an increase in the Fe$^{2+}$ component at 708.7 eV in the Fe$2p_{3/2}$ peak (compare inset in Fig. 5a). Adsorption of 10 L methanol at 280 K has no effect on the Fe$2p$ spectrum for the clean or Fe rich surface (not shown) and also little effect on the O$1s$ region (not shown). In the C$1s$ region (Figure 5b) however, a peak appears at 286.1 eV on the clean surface, related to adsorption on defects. Heating to 300 K decreases this peak's intensity, while it remains at the same position. On the Fe deposited surface (blue), a peak appears at 286.3 eV, which is 50 % larger than the one obtained for the clean Fe$_3$O$_4$(001) surface. Again, heating to 300 K decreases the intensity of the peak, which remains at the same position.

The C$1s$ spectra show a small shift in the binding energies between the 0.3 ML ML Fe-Fe$_3$O$_4$(001) surface and the clean Fe$_3$O$_4$(001) surface. This shift may be related to changes in the valence charge on the carbon due to changes in the electronegativity of the vicinity. Such an interpretation was suggested for similar differences between oxidized and reduced cerium oxide thin films [31], where the peak position on the reduced surface is shifted to higher binding energy. Alternatively, there may be a contribution from molecular methanol adsorbed on Fe adatoms.

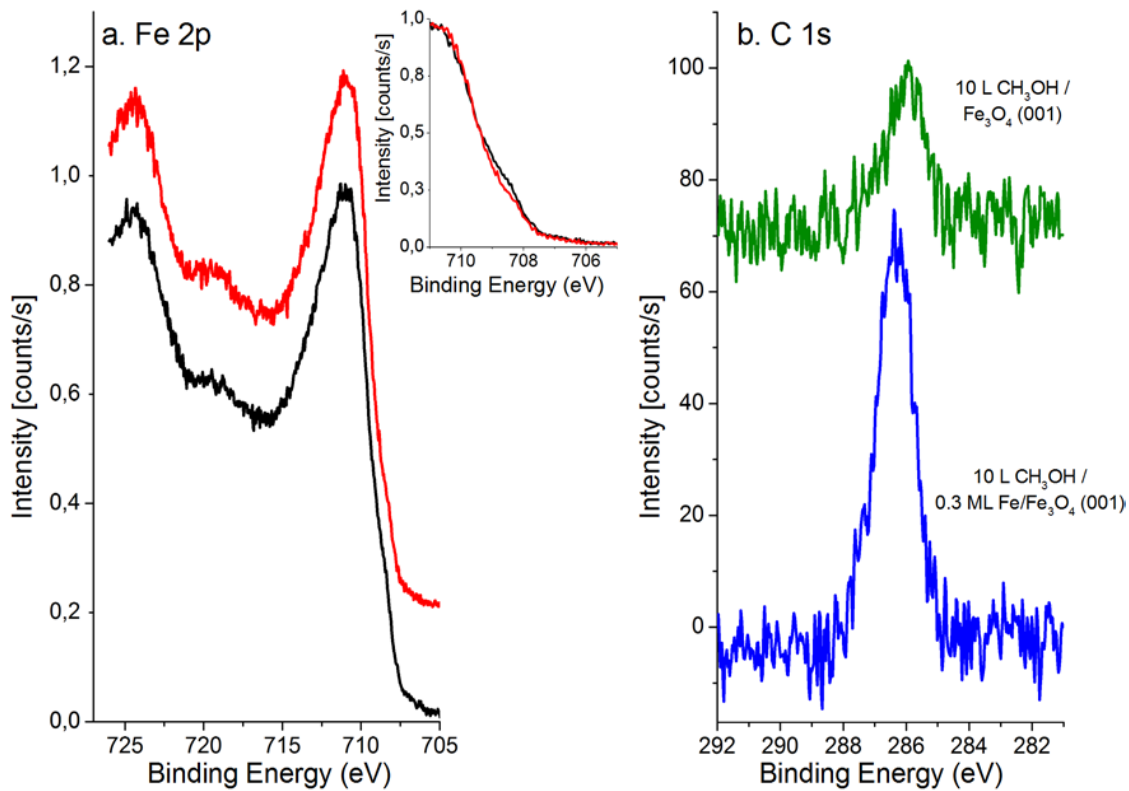

**Figure 5. a.** Fe 2p XPS spectra before (red) and after (black) the deposition of 0.3 ML Fe on the clean Fe$_3$O$_4$(001)-($\sqrt{2} \times \sqrt{2}$)R45° surface at room temperature. b. C 1s XPS spectra from methanol adsorbed at 280 K on the Fe$_3$O$_4$(001)-($\sqrt{2} \times \sqrt{2}$)R45° surface (green line) and the 0.3 ML Fe/Fe$_3$O$_4$(001)-($\sqrt{2} \times \sqrt{2}$)R45° surface (blue line). The data are offset in the y-direction for clarity.

3.4. TPD

Temperature programmed desorption was performed following the adsorption of 10 L CH$_3$OH at 280 K on the clean Fe$_3$O$_4$(001) surface (Figure 6a) and on the 0.3 ML Fe-Fe$_3$O$_4$(001) surface (Figure 6b). Desorption of methanol was monitored by following mass 31 (red dots), which is the most intense cracking fragment of this molecule. By comparing this signal with mass 29 (black dots), possible reaction products can be determined [5]. On the pristine surface, the TPD spectra exhibit two peaks; a sharp peak at 335 K, and a broad shoulder in the region between 400 and 580 K. While the peak at 335 K has a similar shape and intensity for both masses, indicative of molecular methanol, the signal in the region between 400 and 580 K is higher for mass 29. With the addition of 0.3 ML Fe, the low-temperature peak at 335 K slightly shifts to 320 K, and its intensity increases by 42%. As before, the intensity and line shape is similar for both mass 29 and 31. In the region between 400 and 580 K, the peak for mass 31 has its maximum at 450 K, while the peak for mass 29, which is significantly sharper, has the maximum intensity at 470 K. The blue curves in figure 6 result from substracting the smoothed signal for mass 31 from the smoothed mass 29 spectrum. This procedure removes the contribution of methanol from the mass 29 signal, leaving only that of formaldehyde. For the clean $Fe_3O_4(001)$-($\sqrt{2} \times \sqrt{2}$)R45° surface a broad peak is observed with a peak at 480 K, which is increased in intensity following the deposition of 0.3 ML Fe.

The peak at 470 K is most likely attributable to desorption of formaldehyde, as observed previously on $Fe_3O_4(111)$ [14]. The mass spectra show no evidence for $CH_4$ or $C_XH_Y$, which would have been related to C–O bond cleavage [35]. Evolution of CO and $H_2$ was also not observed.

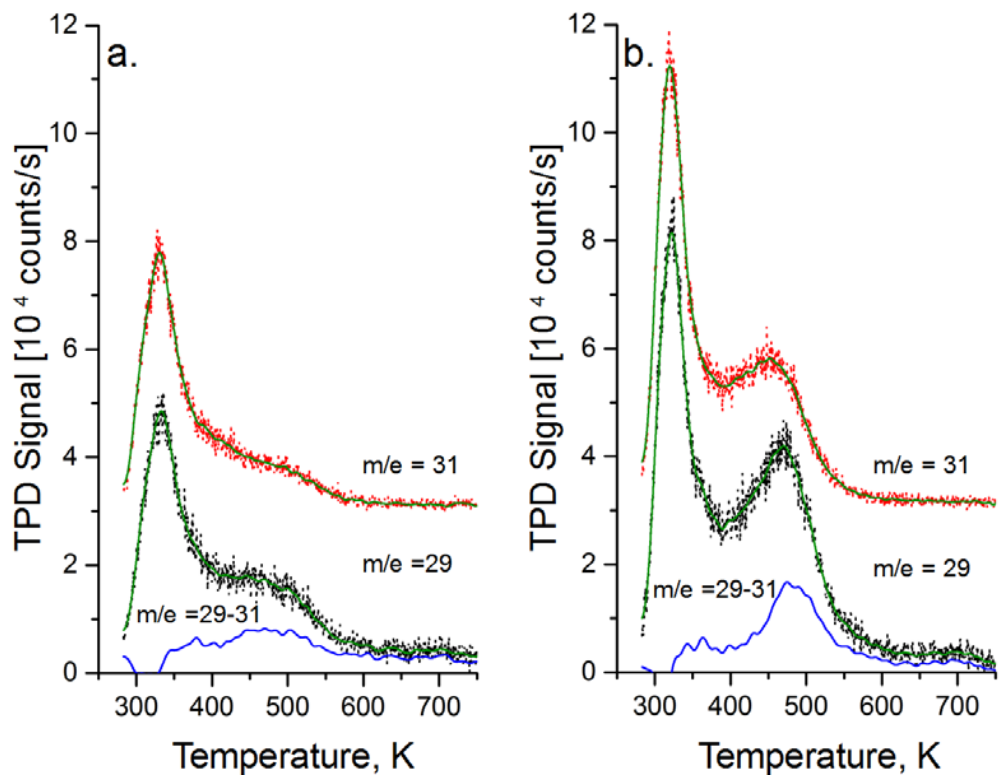

**Figure 6.** TPD spectra of 10L of CH$_3$OH dosed at 280 K. (a) Fe$_3$O$_4$(001)-($\sqrt{2} \times \sqrt{2}$)R45° surface; (b) 0.3 ML Fe - Fe$_3$O$_4$(001)-($\sqrt{2} \times \sqrt{2}$)R45° surface. The data are offset in the y-direction for clarity. The blue curve, calculated by subtracting the smoothed signal from mass 29 from that of the smoothed signal of mass 31, represents the desorption of formaldehyde.

## 4. DISCUSSION

4.1 Adsorption at defects

On the basis of the STM, XPS and TPD results described above it is clear that methanol adsorption is restricted to defect sites on the Fe$_3$O$_4$(001) surface at room temperature. This is in contrast to formic acid, which was previously found to dissociate at regular lattice sites on this surface, resulting in a monolayer of bidentate formate species [17]. That the surface dissociates formic acid suggests that the cation-anion site separation is not prohibitive, but rather the acid-base strength of the surface atoms is insufficient to induce dissociation of weaker acids, such as methanol. On the basis of the STM data the active sites for dissociation are identified as step edges, Fe adatoms, incorporated Fe defects, and APDBs; we discuss these in turn in the following.

The reactivity of step edges on metal oxide surfaces is well documented [4,36,37], and is linked to the coordinative unsaturation of the atoms located there. In the present case, the step edge structures are not definitively known.

Following the procedure of Henrich [20], we have reevaluated the step stability in terms of covalent stability (coordinative unsaturation) for the SCV reconstructed surface. The most stable step parallel to the $Fe_{oct}$ rows (denoted B-α* by Henrich) exposes $Fe_{tet}$ atoms with only one dangling bond each (Figure 7a). All other configurations expose $Fe_{oct}$ atoms with three dangling bonds at each atom, which are likely more reactive. Perpendicular to the rows several different configurations are similarly stable, and all expose both, $Fe_{oct}$ atoms with three dangling bonds per atom, and $Fe_{tet}$ atoms, which have two dangling bonds (one example is shown in Figure 7b). The higher coordinative unsaturation of cations at the perpendicular steps likely makes these sites reactive.

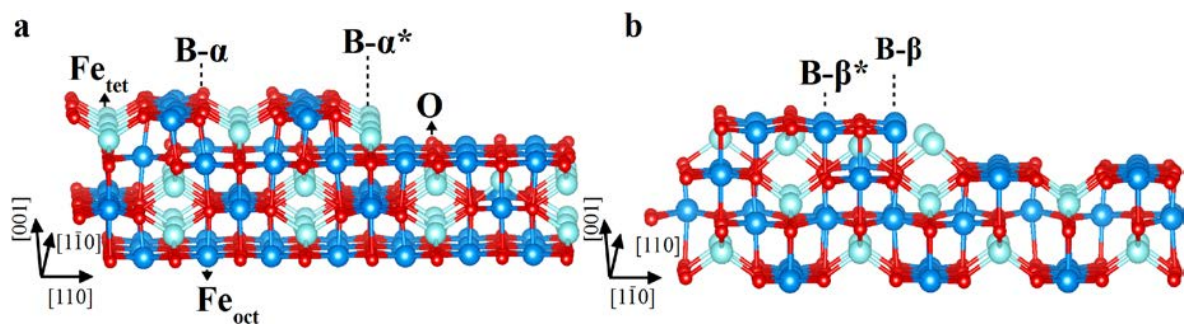

Figure 7. Structural models of step edges for the $Fe_3O_4$(001) surface. (a) Steps parallel to the iron ion rows along [1$\bar{1}$0] (B-α type). (b) Steps perpendicular to the iron ion rows along [110] (B-β type). Labels after ref. [20]

This suggests that the perperndicular steps should be more reactive, as they expose Fe with higher coordinative unsaturation. However, our results showed some degree of reactivity for both kind of steps to methanol adsorption, as obseved in Figure 2 and 3.

Fe adatoms are not a common defect when the $Fe_3O_4$(001) surface is annealed in oxygen, but they become more prevalent when the surface is prepared in reducing conditions [38], when the $Fe_3O_4$ bulk is Fe-rich [39, 40], or when Fe is evaporated onto the surface [24]. The reactivity of the Fe adatoms is again most likely linked to coordinate unsaturation, because such cations have only two bonds to surface oxygen. The fuzzy appearance of the adsorbed methanol (inset Fig. 3), which indicates mobility underneath the STM tip, could mean that the molecule is adsorbed more weakly as compared to the other defects. This is probably because dissociation of the molecule is precluded by the lack of nearby lattice oxygen that can receive the acid proton. Previous experimental and theoretical studies have found that adsorption of H is energetically unfavorable for an $O_{surface}$ with a subsurface $Fe_{tet}$ neighbor [41].

In addition to Fe adatoms, this paper reports a new defect linked to excess Fe that appears in STM as a pair of bright protrusions on opposite $Fe_{oct}$ rows. The density of such defects scales with the amount of deposited Fe, and resembles protrusions observed following incorporation of Ni, Ti and Co atoms in the subsurface vacancies of the SCV reconstruction [28]. By analogy, it is proposed that deposited Fe atoms can enter the subsurface and occupy one of the $Fe_{oct}$ vacancies present in the third layer. This induces the $Fe_{int}$ interstitial to move and occupy the other $Fe_{oct}$, resulting in a structure that locally resembles a bulk-truncated $Fe_3O_4$ lattice. It is interesting that such a defect would be reactive, because the surface layer of the bulk truncated surface should differ little from that of the SCV reconstruction, save for some small relaxations (<0.1 Å). The coordinative unsaturation of the Fe and O atoms are similar, and DFT+U calculations predict that both surfaces contain only $Fe^{3+}$-like cations in the surface layer [16]. A key difference affecting reactivity might be the local electronic structure: Density-of-states plots for the bulk truncation exhibit significantly greater density of (empty) states above the Fermi level than the SCV reconstruction due to $Fe_{oct}^{2+}$-like cations in the third layer [42, 43]. The presence of such states can make the region a stronger Lewis acid site, and more receptive to the electrons from the methoxy.

As discussed above, the APDB is an interruption of the vacancy-interstitial pattern in the second and third layers of the SCV reconstruction, and forms such that four $Fe_{oct}$ cations meet in the third layer at the junction (see Figure 1). Locally, such a configuration also resembles again the unreconstructed lattice. Thus the reactivity can be explained in similar terms to the incorporated-Fe point defect.

4.2. Reaction channels for the adsorption of $CH_3OH$ on $Fe_3O_4$ (001)

TPD analysis shows that for both surfaces, the signals for masses 29 and 31 match perfectly over the range between 280 and 350 K. This is evidence that the peak around 300 K observed is due to desorption of $CH_3OH$ only [5]. Similar observations have been reported on $Fe_3O_4(111)$ [14] and $TiO_2(110)$ [44], where peaks in the same range of desorption temperatures have been assigned to the recombinative desorption of $CH_3OH$. Deposition of Fe results in an increased desorption in this region, which is probably linked to molecular methanol adsorbed at the Fe adatoms.

The increased desorption in the region between 400 and 580 K suggests that the adsorption of methanol increases in the presence of Fe-related defects. Following the behavior of the spectra for both masses, is clear that, in the high-temperature range, the intensity and the shape of both spectra are different; this observation indicates the presence of another species in addition to methanol. Specifically, mass 31 has a contribution from methanol, while mass 29 has contributions from both methanol and formaldehyde ($H_2CO$); the latter is often observed as product of methanol reaction. Consequently, we assign the signal at high temperature as due to partial oxidation of methanol to formaldehyde.

Partial oxidation has been identified on other oxide surfaces as one of the main possible reactions. Interestingly, when the $Fe_3O_4(001)$ surface was modified to have additional Fe adatoms and incorporated Fe defects a similar increase was observed in the peak at 320 K and the high-temperature products linked to the disproportionation reaction (equation 1). Given the relative abundance of Fe adatoms and incorporated Fe defects (approx. 50:50 at 0.3 ML coverage), and the probability that the Fe adatoms cannot dissociate $CH_3OH$ and thus contributes to the lower temperature peak, it seems likely that the increase in formaldehyde production is due to the subsurface Fe defects. This can be because this defect promotes the disproportionation reaction by adsorbing two methoxy species in close proximity. Given the structural and electronic similarities between the incorporated Fe defect and the APDB, this defect also likely promotes the disproportionation reaction.

The temperature for the disproportionation reaction appears to depend on various factors such as the oxidation state of the metal oxide. In the case of vanadium oxide supported on $CeO_2$, formaldehyde desorption curves have shown signals in temperature ranges from 500 to 610 K, depending on the oxidation state of vanadium [45]. At the $CeO_2$ (111) surface, methanol was oxidized to formaldehyde and water at 680 K after that methanol dissociation had occurred at oxygen vacancies [10]. On the other hand, complete dehydrogenation of methanol to CO and $H_2$ has been reported on highly reduced ceria surfaces [9]. It is important to note that a similar chemistry occurs on the $Fe_3O_4$(001) surface despite the lack of oxygen vacancies, and that defects related to excess $Fe^{2+}$ play the important role. Given the available evidence it appears that the adsorption of multiple methoxy species in close proximity at the incorporated Fe defect may promote the disproportionation reaction.

## 5. CONCLUSIONS

We have studied the adsorption of methanol on the $Fe_3O_4$ (001) surface using TPD, STM and XPS. Methanol adsorbs dissociatively on $Fe_3O_4$ (001) at 280 K at defect sites that were identified as step edges, antiphase domain boundaries (APDB), iron adatoms and incorporated-Fe defects. Whereas adsorption at the steps and Fe adatoms can be explained in terms of coordinative unsaturation, reactivity at the APDBs and incorporated Fe defects is linked to the local electronic structure; specifically to the presence of $Fe^{2+}$ cations in the surface layers. We propose that the adsorption of multiple methoxy species at the latter two defects promotes a disproportionation reaction to form methanol and formaldehyde.

**ACKNOWLEDGEMENT.**

G.S.P., R.B., O.G., J.H., and J.P. acknowledge funding from the Austrian Science Fund START prize Y 847-N20 and project number P24925-N20. R.B. and O.G. acknowledge a stipend from the Vienna University of Technology and the Austrian Science Fund as part of the doctoral college SOLIDS4FUN (W1243). U.D. and J.P. acknowledge support by the European Research Council (Advanced Grant "OxideSurfaces"). M.S. was supported by the Austrian Science Fund (FWF) within SFB F45 "FOXSI".